\documentclass[aps,prc,twocolumn,showpacs,floatfix]{revtex4-1}

\usepackage{graphicx}
\usepackage{dcolumn}
\usepackage{bm}
\usepackage{ulem}
\usepackage{color}

\newcommand{\al}{$\alpha$}
\newcommand{\g}{$\gamma$}
\newcommand{\raa}{($\alpha$,$\alpha$)}
\newcommand{\raX}{($\alpha$,$X$)}

\newcommand{\rag}{($\alpha$,$\gamma$)}
\newcommand{\ran}{($\alpha$,n)}

\newcommand{\rna}{(n,$\alpha$)}

\newcommand{\rap}{($\alpha$,p)}
\newcommand{\rapp}{($\alpha$,2p)}

\newcommand{\rpa}{(p,$\alpha$)}

\newcommand{\rpt}{(p,t)}

\newcommand{\stot}{$\sigma_{\rm{reac}}$}

\newcommand{\Nsv}{$N_A$$\left< \sigma v \right>$}

\newcommand{\alppro}{$\alpha p$-process}
\newcommand{\arvi}{$^{36}$Ar}
\newcommand{\arviii}{$^{38}$Ar}
\newcommand{\arnull}{$^{40}$Ar}
\newcommand{\caviii}{$^{38}$Ca}
\newcommand{\caix}{$^{39}$Ca}
\newcommand{\canull}{$^{40}$Ca}
\newcommand{\cai}{$^{41}$Ca}
\newcommand{\caii}{$^{42}$Ca}
\newcommand{\caiii}{$^{43}$Ca}
\newcommand{\caiv}{$^{44}$Ca}
\newcommand{\kix}{$^{39}$K}
\newcommand{\ki}{$^{41}$K}
\newcommand{\kiii}{$^{43}$K}
\newcommand{\sci}{$^{41}$Sc}
\newcommand{\tii}{$^{41}$Ti}
\newcommand{\tiii}{$^{42}$Ti}

\begin{document}

\title{Cross sections of $\alpha$-induced reactions slightly below
  doubly-magic $^{40}$Ca from the statistical model
}

\author{P.~Mohr} \email[Electronic Address: ]{mohr@atomki.mta.hu}
\affiliation{Diakonie-Klinikum, Schw\"{a}bisch Hall D-74523, Germany}
\affiliation{Institute for Nuclear Research (Atomki), Debrecen H-4001, Hungary}
\author{R.~Talwar}
\affiliation{Physics Division, Argonne National Laboratory, Argonne, IL 60439 USA}
\author{M. ~L. ~Avila}
\affiliation{Physics Division, Argonne National Laboratory, Argonne, IL 60439 USA}
\date{\today}

\begin{abstract}
New experimental data for the $^{38}$Ar($\alpha$,n)$^{41}$Ca and
$^{38}$Ar($\alpha$,p)$^{41}$K reactions were used to find a best-fit set of
parameters for statistical model calculations. The very good agreement between
the experimental data and the best-fit calculation confirms the applicability
of the statistical model for nuclei in the vicinity of the doubly-magic
$^{40}$Ca despite of their relatively low level densities. The present study
investigates the sensitivities and finds that the $\alpha$-nucleus and the
nucleon-nucleus potentials are the most important ingredients for the
calculation of ($\alpha$,n) and ($\alpha$,p) reactions in the statistical
model. Furthermore, the width fluctuation correction plays an essential role
in the peculiar case of $^{38}$Ar. The best-fit parameters from $^{38}$Ar are
applied to the mirror nucleus $^{38}$Ca and the neighboring $^{36}$Ar and
$^{40}$Ar nuclei. For $^{38}$Ca this results in an astrophysical reaction rate
of the $^{38}$Ca($\alpha$,p)$^{41}$Sc reaction which has a flatter temperature
dependence compared to all previous calculations. For $^{40}$Ar a better
reproduction of $^{40}$Ar($\alpha$,p)$^{43}$K data from literature is
obtained. The disagreement between calculation and an early experimental data
point for the $^{36}$Ar($\alpha$,p)$^{39}$K reaction persists.
\end{abstract}

\maketitle

\section{Introduction}
\label{Sec:intro}
The cross sections and stellar reaction rates of \al -induced reactions play
an important role in various astrophysical scenarios. For targets slightly
below the doubly-magic \canull , the \arviii \rap \ki\ and \arviii \ran
\cai\ reactions have been studied recently because their reverse \ki \rpa
\arviii\ and \cai \rna \arviii\ reactions have been identified to affect the
abundance of the relatively short-lived \cai\ nucleus in the early solar
system. The dedicated experiment has provided several \rap\ and \ran\ cross
sections at low energies (see \cite{Tal18} and references
therein). Furthermore, \rap\ reactions on isospin $T_z = -1$ nuclei like
$^{30}$S \cite{Kahl18,Long18}, $^{34}$Ar \cite{Long17,Schmidt17}, and
$^{38}$Ca play a key role in the so-called \alppro\ in X-ray bursters
\cite{Wall81,Schatz06,Par13}. Obviously, direct experiments on these
short-lived $T_z = -1$ nuclei are very difficult, and often indirect
information from \rpt\ reactions is used to estimate the stellar reaction
rates \cite{Long17,Long18,OBri09}.

In general, a very reasonable description of \al -induced reaction cross
sections at low energies has been found for nuclei in the $A \approx 20 - 50$
mass range \cite{Mohr15}. It is based on the statistical model (StM) in
combination with the widely used simple 4-parameter \al -nucleus optical model
potential (A-OMP) by McFadden and Satchler \cite{McF66}. Interestingly,
further ingredients of the StM play a very minor role in this mass range
because the A-OMP defines the total \al -induced reaction cross section
\stot\ which is typically dominated by either the \ran\ or the
\rap\ channel. However, the simple approach of a dominating \ran\ or
\rap\ channel does not hold in vicinity of the doubly-magic \canull . The
$Q$-values of the \arviii \ran \cai\ and \arviii \rap \ki\ reactions are both
significantly negative, and thus the description of these reactions in the StM
requires additional care. 

It is the scope of the present paper to present the relevant details and
improvements of the StM calculations in \cite{Tal18} for the \arviii \ran
\cai\ and \arviii \rap \ki\ reactions. In addition, these improvements will be
used to predict \al -induced cross sections for the mirror nucleus
\caviii\ and for \arvi\ and \arnull . For the latter two argon isotopes
significant discrepancies between the StM calculations and very old
experimental data \cite{Schwartz56} were identified in \cite{Mohr15}. Note
that new experimental data for the two further outliers in \cite{Mohr15},
$^{23}$Na \cite{Avila16,How15,Tom15} and $^{33}$S \cite{And17} (see also
\cite{Mohr14}), supersede previous data \cite{Alm14,Bowers13} and are now in
better agreement with the earlier predictions in \cite{Mohr15}.

\section{\al -induced cross sections in the statistical model for targets
  close to $^{40}$Ca}
\label{sec:theo}

\subsection{General remarks}
\label{sec:gen}
The cross sections of \al -induced reactions in the present study have been
calculated within the StM. By definition, the StM provides average cross
sections which are based on the assumption of a sufficiently high level
density in the compound nucleus. In the following, the compound nucleus
\caii\ and the system \arviii $+$\al\ with the \arviii \ran \cai\ and \arviii
\rap \ki\ reactions have been chosen as an example.

In reality, the \ran\ and \rap\ cross sections for targets close to
\canull\ are composed of the contributions of several resonances which may be
broad and overlapping. A typical experiment averages these resonance
contributions over an energy interval $\Delta E$ which is essentially defined
by the energy distribution of the beam and the energy loss of projectiles in
the target:
\begin{equation}
\Delta E = E^{\rm{exp}}_{\rm{max}} - E^{\rm{exp}}_{\rm{min}}
\label{eq:DE}
\end{equation}
with $E^{\rm{exp}}_{\rm{max}}$ and $E^{\rm{exp}}_{\rm{min}}$ being the highest
and lowest experimental energy (given as $E_{\rm{c.m.}}$ in the center-of-mass
system). This experimental energy window $\Delta E$ corresponds to a window
$\Delta E^\ast$ in excitation energies $E^\ast$ in the compound nucleus from
$E^\ast_{\rm{min}} = Q_\alpha + E^{\rm{exp}}_{\rm{min}}$ to $E^\ast_{\rm{max}}
= Q_\alpha + E^{\rm{exp}}_{\rm{max}}$ with the $Q$-value $Q_\alpha$ of the
\rag\ reaction.

Depending on the experimental conditions, $\Delta E$ may be of the
order of a few keV (for primary beams and thin targets) or much larger
(typically a few hundred keV for secondary and/or radioactive ion beams with
low intensities and the required thick targets). Obviously, for a successful
application of the StM a sufficient number of resonances has to be located
within the experimental energy interval $\Delta E$. Otherwise, the StM is only
able to provide the average trend of the experimental data.

Besides theoretical estimates from level density formulae, there is a simple
experimental criterion for the applicability of the StM. As long as the
excitation functions of the \ran\ and \rap\ reactions show a relatively smooth
energy dependence, the application of the StM should be justified. Contrary,
the StM must fail to reproduce experimental excitation functions where the data
points show significant scatter from the contributions of individual
resonances. The new experimental data for the chosen examples \arviii \ran
\cai\ and \arviii \rap \ki\ \cite{Tal18} show a relatively smooth energy
dependence (except the two lowest data points of the \rap\ reaction), and thus
the StM should be applicable in the present case although the level densities
in the semi-magic \arviii\ ($N = 20$) target nucleus and \caii\ ($Z = 20$)
compound nucleus remain relatively small.

\subsection{Formalism of the statistical model}
\label{sec:form}
The Hauser-Feshbach StM \cite{Hau52} is described in many publications. A
detailed description for its application to low-energy reactions and the
calculation of astrophysical reaction rates is provided e.g.\ in
\cite{Rau11}. Here we briefly repeat the essential definitions which are
relevant in the following discussion.

In a schematic notation the reaction cross section in the Hauser-Feshbach (HF)
StM \cite{Hau52} is proportional to 
\begin{equation}
\sigma(\alpha,X)_{\rm{HF}} \sim \frac{T_{\alpha,0} T_X}{\sum_i T_i} = T_{\alpha,0}
\times b_X
\label{eq:HF}
\end{equation}
with the transmission coefficients $T_i$ into the $i$-th open channel and the
branching ratio $b_X = T_X / \sum_i T_i$ for the decay into the channel
$X$. The total transmission is given by the sum over all contributing
channels: $T_{\rm{tot}} = \sum_i T_i$. The $T_i$ are calculated from optical
potentials for the particle channels and from the gamma-ray strength function
for the photon channel. The $T_i$ include contributions of all final
states $j$ in the respective residual nucleus in the $i$-th exit channel. In
practice, the sum over all final states $j$ is approximated by the sum over
low-lying excited states up to a certain excitation energy $E_{\rm{LD}}$
(these levels are typically known from experiment) plus an integration over
a theoretical level density for the contribution of higher-lying excited states:
\begin{equation}
T_i = \sum_j T_{i,j} \approx 
\sum_j^{E_j < E_{\rm{LD}}} T_{i,j} +
\int_{E_{\rm{LD}}}^{E_{\rm{max}}} \rho(E) \, T_i(E) \, dE
\label{eq:Ti}
\end{equation}
$T_{\alpha,0}$ in Eq.~(\ref{eq:HF}) refers to the entrance channel with the
target nucleus (\arviii\ in the present example) in the ground state and
defines the total \al -induced reaction cross section \stot . (For the
calculation of astrophysical reaction rates, thermally excited states in the
target have to be considered in addition.)

There are correlations between the incident and outgoing waves which have to
be taken into account by a so-called width fluctuation correction factor
(WFCF) $W_{\alpha X}$: 
\begin{equation}
\sigma(\alpha,X) = \sigma(\alpha,X)_{\rm{HF}} \times W_{\alpha X}
\label{eq:StM}
\end{equation}
The WFCFs approach unity at higher energies as soon as many reaction channels
are open. At low energies the WFCFs lead to an enhancement of the
compound-elastic cross section and to a reduction of the \raX\ reaction cross
sections. Several methods have been suggested to calculate these WFCFs (see
further discussion below).

\subsection{Ingredients of the statistical model and sensitivities}
\label{sec:sens}
From Eqs.~(\ref{eq:HF})--(\ref{eq:StM}) it is obvious that the calculated cross
sections in the StM depend on the transmissions $T_i$ which in turn depend on
the following ingredients. The neutron and proton transmissions $T_n$ and
$T_p$ are calculated from nucleon optical model potentials (N-OMP); the
\al\ transmission $T_\alpha$ depends on the chosen \al -nucleus optical model
potential (A-OMP), and the \g\ transmission $T_\gamma$ is given by the \g -ray
strength function (GSF). Note that other channels are typically closed at low
energies (or have only very minor contributions).

All transmissions $T_n$, $T_p$, $T_\alpha$, and $T_\gamma$ have a further
implicit dependence on the chosen level density (LD) which results from
Eq.~(\ref{eq:Ti}); $T_{\alpha,0}$ in Eq.~(\ref{eq:HF}) is independent of the
chosen LD. For completeness we point out that the choice of a LD in
Eq.~(\ref{eq:Ti}) should not be confused with the required sufficiently high
LD in the energy interval $\Delta E$ in Eq.~(\ref{eq:DE}). The former is a
choice for a calculation; the latter is the basic prerequisite for the
applicability of the StM; as such, it is a physical property of the system
under investigation and cannot be chosen or even changed in calculations.

Summarizing, the cross section in the StM depends explicitly on the chosen
A-OMP, N-OMP, and GSF, and implicitly on the chosen LD. As will be shown, the
A-OMP is the most important parameter, whereas the remaining parameters N-OMP,
GSF, and LD have relatively minor influence on the \ran\ and \rap\ cross
sections. Typically, the \rag\ cross section is sensitive to a combination of
all parameters as soon as the energy exceeds the \ran\ or \rap\ threshold.

In recent work, two different approaches have been followed to study the
sensitivities of the calculated cross sections in the StM. A strictly
mathematical definition for the sensitivity is for example provided in
\cite{Rau12}. Eq.~(1) of \cite{Rau12} defines a relative sensitivity of 1.0 if
a variation of the input parameter (typically, a transmission $T_i$) by a
certain factor (e.g., a factor of two) leads to a variation of the resulting
cross section by the same factor. A more empirical approach was followed in
\cite{Tal18,Mohr17}. Here a reasonable variation of the input parameters $T_i$
was estimated from the choice of different parametrizations (e.g., a
reasonable variation of $T_\alpha$ was estimated from the choice of different
A-OMPs), and finally a $\chi^2$-based assessment was used to select
combinations of A-OMPs, N-OMPs, GSFs, and LDs. These $\chi^2$-selected
combinations are able to reproduce the available experimental data within the
measured energy range and should be used for the prediction of cross sections
outside the measured energy range with improved reliability. Although the
method of both approaches is different, the conclusions for the reactions
under study in the present work are practically identical.

The last ingredient of the StM calculations is the chosen method for the
calculation of the WFCFs in Eq.~(\ref{eq:StM}). Typically, for \al -induced
reactions the importance of the WFCFs is very minor. However, this typical
behavior does not apply for the \al -induced reactions on \arviii , and thus
the WFCFs have to be taken into account for \arviii\ and neighboring target
nuclei. 

As studied in detail in the mass range $20 \le A \le 50$, a more or less
generic behavior of \al -induced reaction cross sections is found
\cite{Mohr15}. As soon as either the \ran\ or \rap\ particle channel is open,
this channel typically dominates: $T_{\alpha,0} \ll T_n$ or $T_p$, and thus
the branchings $b_n$ or $b_p$ approach unity, see Eq.~(\ref{eq:HF}). This is
obvious for the neutron channel because of the missing Coulomb barrier; but it
holds also for the proton channel because the Coulomb barrier in the proton
channel is much lower than in the \al\ channel. Consequently, either the
\rap\ or \ran\ channel contributes with typically 90\% or more to
\stot\ \cite{Mohr15}. Under these conditions the WFCFs for the \raX\ channels
become negligible because even a dramatic enhancement of the weak
compound-elastic channel, e.g.\ say $W_{\alpha \alpha} = 2$, does practically
not affect and reduce the dominating \rap\ or \ran\ cross sections.

For \arviii\ the $Q$-values for the typically dominating \ran\ or
\rap\ channels are both significantly negative ($Q_{\rm{n}} = -5.22$ MeV and
$Q_{\rm{p}} = - 4.02$ MeV). As a consequence, the usual approximation
$T_{\alpha,0} \ll T_n$ or $T_p$ does not hold for \arviii , and after
formation of the compound nucleus \caii , it may also decay back to the
\al\ channel (compound-elastic channel). As the WFCFs enhance this channel and
reduce the \ran\ and \rap\ channels, it is important to study different
methods for the calculation of the WFCFs.

The results in the following sections are based on calculations with the
widely used code TALYS, version 1.80, \cite{TALYS} which provides the choice
of different A-OMPs, N-OMPs, GSFs, LDs, and methods for the calculation of
WFCFs. In addition, the code has been modified to implement the recently
suggested A-OMP ATOMKI-V1 for heavy targets ($A \gtrsim 90$) \cite{Mohr13}. In
Sect.~\ref{sec:ar38} a detailed study of the sensitivies is provided for the
target nucleus \arviii . The results for \arviii\ are used to constrain the
parameters for the mirror nucleus \caviii\ and for the neighboring
isotopes \arvi\ and \arnull\ in Sect.~\ref{sec:other}.

\section{Results for $^{38}$Ar}
\label{sec:ar38}
A $\chi^2$ search has been performed to find the best combination of input
parameters for the new \arviii \ran \cai\ and \arviii \rap \ki\ data
\cite{Tal18}. As already summarized in \cite{Tal18}, the best-fit parameters
consist of the A-OMP by McFadden and Satchler \cite{McF66} and the
TALYS-default N-OMP by Koning and Delaroche \cite{Kon03}. The $\chi^2$ search
is not very sensitive to the chosen LD, but the smallest $\chi^2$ is found for
the LD calculated from the generalized superfluid model
\cite{Ign79,Ign93}. Finally, the \ran\ and \rap\ cross sections are
practically insensitive to the choice of the GSF, leading to a $\chi^2$ per
experimental data point between 4.76 and 4.78 and an average deviation of
13.6\% to 13.7\% for the best-fit A-OMP, N-OMP, and LD, and an arbitrary choice
of the GSF. This minor sensitivity to $T_\gamma$ and the GSF is obvious from
Eq.~(\ref{eq:HF}) where $T_\gamma$ appears only as a minor contribution to
the sum $\sum_i T_i$ in the denominator. Consequently, the following discussion
provides detailed information on the sensitivities of 
the WFCFs (Sect.~\ref{sec:WFCF}), 
the A-OMPs (Sect.~\ref{sec:AOMP}), 
the N-OMPs (Sect.~\ref{sec:NOMP}), 
and the LDs (Sect.~\ref{sec:LD}), 
whereas a discussion of GSFs is omitted. The best-fit parameters (as stated
above) will be used as a reference in the following presentation.

\begin{figure}[htb]
\includegraphics[width=\columnwidth,clip=]{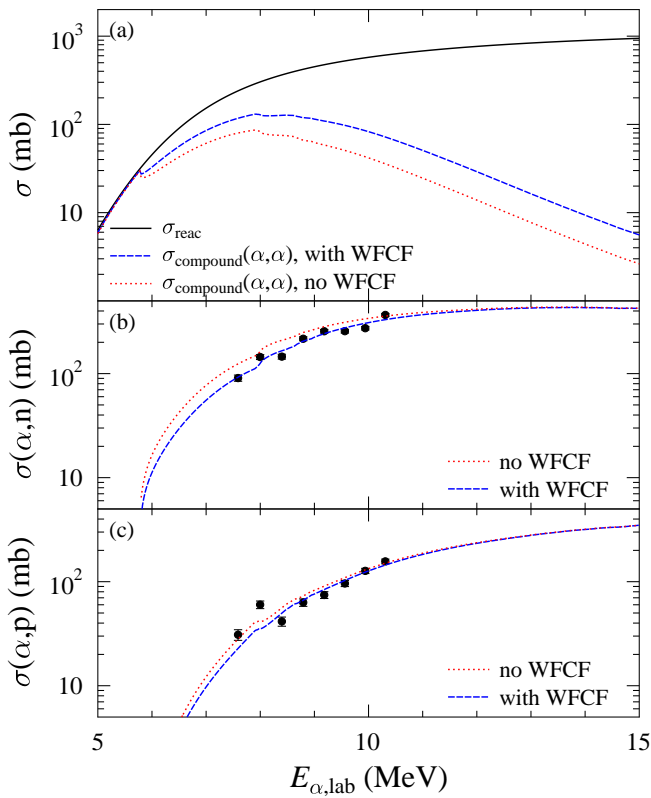}
\caption{
\label{fig:ar38_summary}
(Color online)
Total cross section \stot\ and compound-elastic
$\sigma_{\rm{compound}}$\raa\ for \arviii\ (upper, a), calculated without WFCF
(dotted red) and with WFCF (dashed blue). Because of the increased
$\sigma_{\rm{compound}}$\raa , the \ran\ (middle, b) and \rap\ (lower, c)
cross sections are reduced by the WFCF. Further discussion see text.
}
\end{figure}

Fig.~\ref{fig:ar38_summary} illustrates the peculiar behavior of \arviii\ and
the importance of the WFCF. The upper part (a) of Fig.~\ref{fig:ar38_summary}
shows the total cross section \stot\ from the reference calculation and the
compound-elastic $\sigma_{\rm{compound}}$\raa . Below the \ran\ and
\rap\ thresholds, \stot\ is dominated by the compound-elastic
contribution. And even up to almost 10 MeV, there is a significant
compound-elastic contribution (blue dashed line) which is enhanced by the WFCF
(in comparison to the calculation without WFCF, red dotted line). Note that
the data for \arviii\ are shown as a function of $E_{\alpha,{\rm{lab}}}$
whereas the experiment \cite{Tal18} was performed in inverse kinematics.

The \arviii \ran \cai\ and \arviii \rap \ki\ cross sections are shown in the
middle (b) and lower (c) parts of Fig.~\ref{fig:ar38_summary}. Obviously, the
reference calculation (including the WFCF) reproduces the new experimental
data \cite{Tal18} very well, whereas a calculation without WFCF overestimates
the \ran\ and \rap\ cross sections. Of course, the WFCF becomes most relevant
for the energy range where the compound-elastic cross section has a
significant contribution to the total reaction cross section \stot\ (i.e.,
below about 10 MeV). At the highest energies under study around 15 MeV, there
is still a significant enhancement of $\sigma_{\rm{compound}}$\raa\ by about a
factor of two from the WFCF; however, here $\sigma_{\rm{compound}}$\raa\ has
only a very minor contribution of less than 1\% to the total reaction cross
section \stot , and thus even a significant enhancement of
$\sigma_{\rm{compound}}$\raa\ does practically not affect the dominating
\ran\ and \rap\ channels.

Next, the sensitivities to the different ingredients of the StM will be
studied in detail for the \arviii \ran \cai\ and \arviii \rap
\ki\ reactions. For better visibility, all plots will be normalized to the
reference calculation with the smallest $\chi^2$ (as defined above), and the
same logarithmic scale is chosen for all plots of the ratio $r_{\rm{calc}} =
\sigma_{\rm{mod}}/\sigma_{\rm{ref}}$ between the cross section with a modified
parameter $\sigma_{\rm{mod}}$ and the reference cross section
$\sigma_{\rm{ref}}$. In addition, also the new experimental data \cite{Tal18}
are shown as ratio $r_{\rm{exp}} = \sigma_{\rm{exp}}/\sigma_{\rm{ref}}$. All
Figs.~\ref{fig:ar38_wfcf} $-$ \ref{fig:ar38_ld} with the calculated
sensitivities use the same scale with ratios $r$ between 0.3 and 3.0; only in
Fig.~\ref{fig:ar38_aomp} the vertical size has been increased for better
visibility because of the larger number of lines.

\subsection{Width fluctuation correction factors}
\label{sec:WFCF}
\begin{figure}[htb]
\includegraphics[width=\columnwidth,clip=]{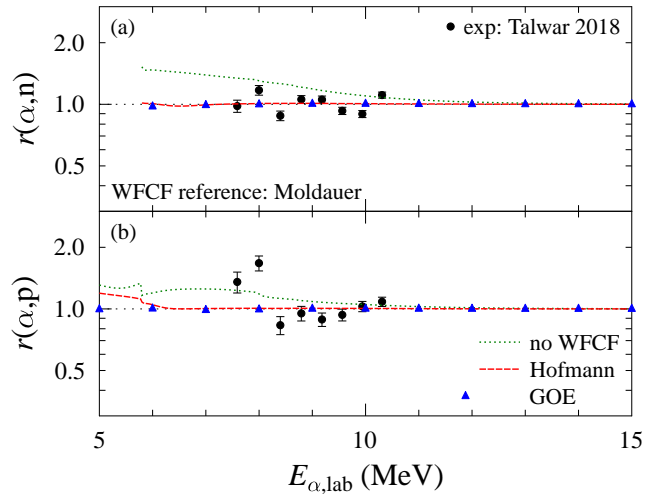}
\caption{
\label{fig:ar38_wfcf}
(Color online) Sensitivity of the \arviii \ran \cai\ and \arviii \rap
\ki\ reactions for different choices for the width fluctuation correction. All
cross sections are normalized to the reference calculation (see text) which
uses the Moldauer approach for the WFCFs. The deviation of the Hofmann
approach (red dashed) is practically negligible, and also the GOE approach
(blue points) does not show a major deviation from the reference
calculation. However, the cross sections without width fluctuation correction
are significantly higher below 10 MeV (green dash-dotted), in particular for
the \ran\ channel.
}
\end{figure}

The WFCFs can be calculated in TALYS from three different models. Widely used
are the approaches by Moldauer \cite{Mol76} and the iterative method by
Hofmann {\it et al.}\ \cite{Hof75}. A more fundamental approach is based on
the Gaussian orthogonal ensemble (GOE) of Hamiltonian matrices \cite{Ver85};
however, in practice this approach requires the calculation of triple
integrals which leads to long computation times in particular at higher
energies. It has been shown for neutron-induced reactions that the simpler
Moldauer approach leads to almost identical results as the elaborate GOE
approach \cite{Hil03,Kaw16}. This has been verified for the \al -induced
reactions on \arviii\ in this study in coarse 1 MeV steps from 3 to 15
MeV. The difference between the Moldauer approach and the Hofmann approach
remains small (see Fig.~\ref{fig:ar38_wfcf}). Therefore, the TALYS default
option by Moldauer was used for the reference calculation.

For completeness it has to be mentioned that the TALYS default setting for the
WFCFs is only active at low energies. Above the separation energy of the
projectile from the target, the WFCFs are assumed unity. Thus, for \al
-induced reactions on \arviii\ with $S_\alpha($\arviii $) = 7.21$ MeV, the
width fluctuation correction is turned off by default at
$E_{\alpha,{\rm{lab}}} = 7.21$ MeV. It is obvious from
Fig.~\ref{fig:ar38_summary} that this is not appropriate for the particular
case of \arviii\ although this TALYS default setting is good for most other
\al -induced reactions. As a consequence, a TALYS calculation for \arviii\ $+$
\al\ with the default settings for the width fluctuation correction shows an
unphysical kink at $E_{\alpha,{\rm{lab}}} = 7.21$ MeV ($E_{\rm{c.m.}} = 6.5$
MeV), see Fig.~5 of \cite{Tal18}.

\subsection{$\alpha$-nucleus optical model potential}
\label{sec:AOMP}
\begin{figure}[htb]
\includegraphics[width=\columnwidth,clip=]{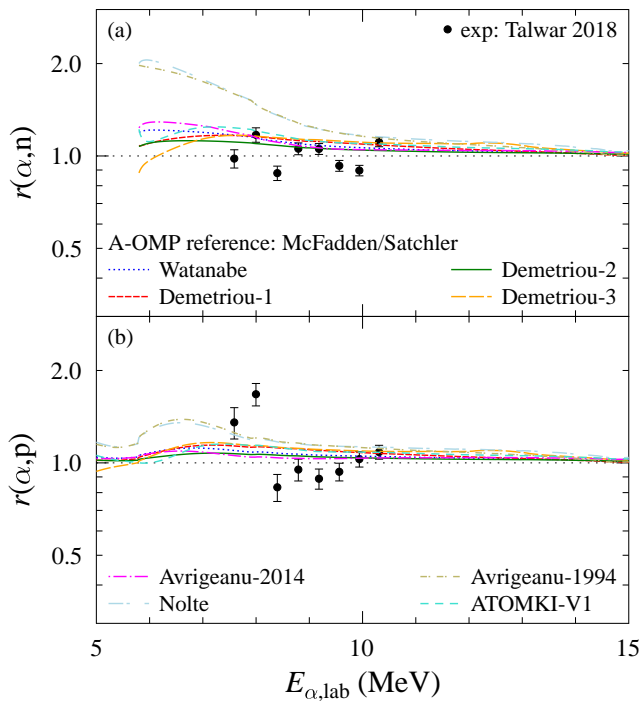}
\caption{
\label{fig:ar38_aomp}
(Color online) Same as Fig.~\ref{fig:ar38_wfcf} for the sensitivity of the
\arviii \ran \cai\ and \arviii \rap \ki\ reactions on the chosen A-OMP. Except
the early potentials by Nolte {\it et al.}\ \cite{Nol87} and Avrigenau {\it et
  al.}\ \cite{Avr94}, the reproduction of the experimental data is quite
good. But clearly the best description is obtained from the McFadden/Satchler
potential \cite{McF66}. 
}
\end{figure}

The A-OMP is the essential ingredient for the calculation of \al -induced
reaction cross sections. It defines the transmission $T_{\alpha,0}$ in the
entrance channel in Eq.~(\ref{eq:HF}) which corresponds to the total reaction
cross section \stot . There are 8 built-in options for the A-OMP in TALYS, and
the recent ATOMKI-V1 potential has been implemented in addition. The results
from the different AOMPs are shown in Fig.~\ref{fig:ar38_aomp}.

As explained in \cite{Mohr13b}, the calculated cross sections from
different A-OMPs are very close to each other at higher energies. Differences
become visible at lower energies below 10 MeV. The potentials by Nolte {\it et
  al.}\ \cite{Nol87} and the earlier version of Avrigeanu {\it et
  al.}\ \cite{Avr94} clearly overestimate in particular the \ran\ channel. As
these potentials have been adjusted at higher energies, such a discrepancy at
low energies is not surprising. Consequently, these potentials should not be
used for the calculation of astrophysically relevant cross sections and
reaction rates.

The potentials by Watanabe \cite{Wat58} (early TALYS default), Demetriou {\it
  et al.}\ \cite{Dem02} in three different versions, Avrigeanu {\it et
  al.}\ \cite{Avr14} in its recent version (new TALYS default), and ATOMKI-V1
\cite{Mohr13} lead to \ran\ and \rap\ cross sections which remain relatively
close to the reference calculation which is based on the simple 4-parameter
potential by McFadden and Satchler \cite{McF66}. However, contrary to the
McFadden/Satchler potential, the other potentials show a trend to overestimate
the \ran\ and \rap\ cross sections by about 10\% to 30\%, leading to a
significantly worse $\chi^2$ for the comparison with the new experimental data
\cite{Tal18}.

The present data confirm the general finding of \cite{Mohr15} that the simple
McFadden/Satchler potential does an excellent job at low energies in the $A
\approx 20 - 50$ mass range. Furthermore, the variation of the calculated
\ran\ and \rap\ cross sections from different A-OMPs is not as dramatic as for
heavy target nuclei with masses above $A \approx 100$ where discrepancies
exceeding one order of magnitude have been seen (e.g.\ \cite{Som98}).

The calculation of astrophysical reaction rates for the \arviii \ran \cai\ and
\arviii \rap \ki\ reactions is further hampered by the negative $Q$-value of
both reactions, leading to numerical complications. This was already discussed
in detail in \cite{Tal18}, and it was concluded that the astrophysical
reaction rate of both reactions has uncertainties which do not exceed a factor
of two for all relevant temperatures. In the most relevant temperature range
around $T_9 \approx 1$ (where $T_9$ is the temperature in Giga-Kelvin) the
uncertainty of the reaction rates is about 30\%.

\subsection{Nucleon-nucleus optical model potential}
\label{sec:NOMP}
\begin{figure}[htb]
\includegraphics[width=\columnwidth,clip=]{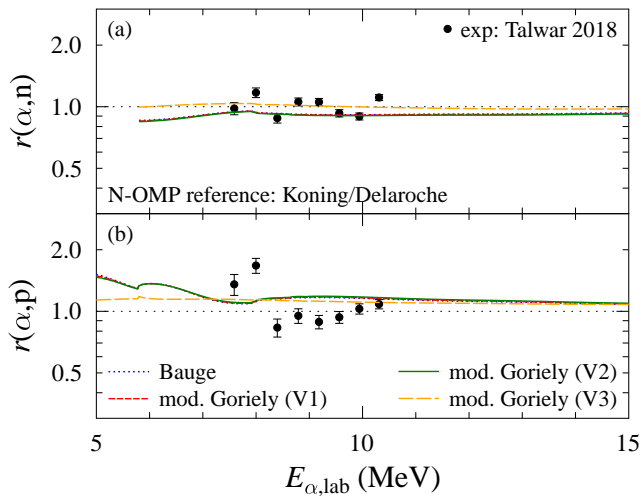}
\caption{
\label{fig:ar38_nomp}
(Color online) Same as Fig.~\ref{fig:ar38_wfcf} for the sensitivity of the
\arviii \ran \cai\ and \arviii \rap \ki\ reactions on the chosen N-OMP. The
JLM-type potentials show a trend to overestimate the \rap\ data and
underestimate the \ran\ data.
}
\end{figure}

The N-OMP essentially defines the branching ratio between the \ran\ and the
\rap\ channel. However, the variation of the \ran\ and \rap\ cross sections
remains relatively small because the major influence on the branching between
\ran\ and \rap\ results from the available phase space. The results for the
different N-OMPs under study are shown in Fig.~\ref{fig:ar38_nomp}.

The best description of the new experimental data is achieved by the TALYS
default N-OMP by Koning and Delaroche \cite{Kon03}. Further N-OMPs in TALYS
are based on the work of Jeukenne, Leujenne, and Mahaux (JLM) \cite{Jeu77} in
the version of Bauge {\it et al.}\ \cite{Bau01}. In addition to the original
Bauge {\it et al.}\ potential, three modifications of the imaginary part of
the JLM-type potential can be selected as suggested by Goriely and Delaroche
\cite{Gor07}.

In general, compared to the default Koning/Delaroche potential, the JLM-type
potentials overestimate the \rap\ channel and underestimate the
\ran\ channel. An exception is the third modification of the JLM-type
potentials where the imaginary strength is increased by a factor of two
(so-called ``{\it{jlmmode 3}}''). This increased imaginary potential favors
the nucleon channels and reduces the compound-elastic contribution, leading to
an overestimation of the \rap\ channel and a good description of the
\ran\ channel.

\subsection{Level density}
\label{sec:LD}
\begin{figure}[htb]
\includegraphics[width=\columnwidth,clip=]{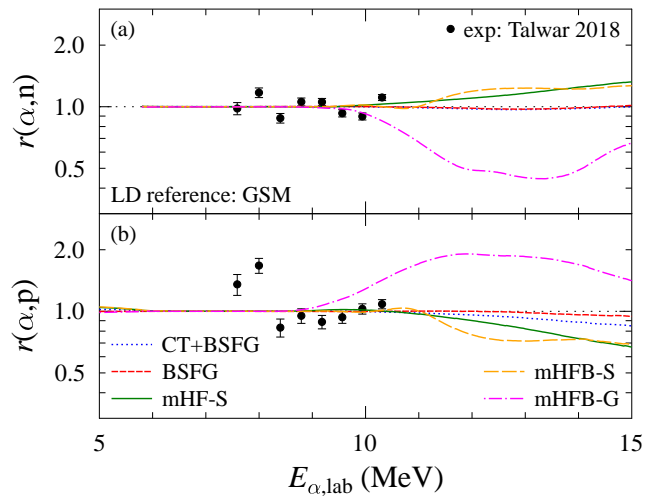}
\caption{
\label{fig:ar38_ld}
(Color online) Same as Fig.~\ref{fig:ar38_wfcf} for the sensitivity of the
\arviii \ran \cai\ and \arviii \rap \ki\ reactions on the chosen LD. The
best-fit is obtained from the generalized superfluid model. At higher energies
above 10 MeV, the mHFB-G LD which is based on Gogny forces, shows a much
larger \rap\ cross section and lower \ran\ cross section than all other
LDs. As expected, at low energies the role of the LD is very minor (see text).
}
\end{figure}

As expected from the discussion around Eq.~(\ref{eq:Ti}), the influence of the
chosen LD on the calculated \ran\ and \rap\ cross sections is almost
negligible at low energies. At these low energies all relevant levels in the
residual nuclei are taken into account explicitly in the
calculations. However, at higher energies the importance of the LD becomes
visible (see Fig.~\ref{fig:ar38_ld}).

The microscopic level density, calculated from a Gogny force \cite{Hil12},
predicts much lower \ran\ cross sections above 10 MeV and higher \rap\ cross
sections. The other available options behave close to the best-fit LD which is
based on the generalized superfluid model \cite{Ign79,Ign93} (labeled ``GSM'')
with a trend of slightly increased \ran\ and slightly decreased \rap\ cross
sections. The other options are labeled by ``CT+BSFG'' for the
constant-temperature model which is matched to the back-shifted Fermi gas
model \cite{Gil65}, ``BSFG'' for the back-shifted Fermi gas model
\cite{Gil65,Dilg73}, ``mHF-S'' for microscopic Hartree-Fock using Skyrme
forces \cite{Gor01}, ``mHFB-S'' for microscopic Hartree-Fock-Bogoliubov using
Skyrme forces \cite{Gor08}, and ``mHFB-G'' for microscopic
Hartree-Fock-Bogoliubov using Gogny forces \cite{Hil12}.

\subsection{Discussion}
\label{sec:disc}
The calculation with the reference parameters is able to reproduce the new
experimental data for the \arviii \ran \cai\ and \arviii \rap \ki\ reactions
with a $\chi^2 \approx 4.8$ (per point) and an average deviation of about
14\%. Obviously, it is not possible to reproduce the two \rap\ cross sections
at the lowest energies which show significant enhancement over the otherwise
smooth energy dependence. This enhancement most likely results from a resonant
contribution which should be located around $E^\ast \approx 13.5$ MeV in
\caii\ with small $J^\pi$ because of the enhanced decay to the \rap\ channel,
i.e.\ towards \ki\ (with low-$J$ states at low excitation energies) and
suppressed \ran\ contribution, i.e.\ towards \cai\ with $J^\pi_{\rm{g.s.}} =
7/2^-$.

The adjustment of the reference parameters via a strict $\chi^2$ assessment
clearly favors the A-OMP by McFadden and Satchler \cite{McF66} in combination
with the default N-OMP by Koning and Delaroche \cite{Kon03}. All other
combinations of A-OMPs and N-OMPs lead to an increased $\chi^2$ per point by
at least 1.1 from its minimum value of about 4.8 to 5.9 and above. Contrary to
this, the LD and the GSF are not well constrained by the new experimental
data. For a well-defined choice of the LD, the available experimental data
should be extended towards higher energies. The GSF could be best constrained
by a measurement of the \arviii \rag \caii\ cross section over a wide energy
range.

In a next step, the reference parameters can be used to calculate \al -induced
cross sections for the mirror target nucleus \caviii\ and the neighboring
argon isotopes \arvi\ and \arnull . As the essential parameters of the StM
have been adjusted to experimental data for \arviii , these calculations
should be more reliable than earlier estimates from global parameter
sets. Furthermore, the relevance of the width fluctuation correction will also
be investigated for \caviii , \arvi , and \arnull .

\section{Results for $^{38}$Ca, $^{36}$Ar, and $^{40}$Ar}
\label{sec:other}

\subsection{$^{38}$Ca}
\label{sec:ca38}
\begin{figure}[htb]
\includegraphics[width=\columnwidth,clip=]{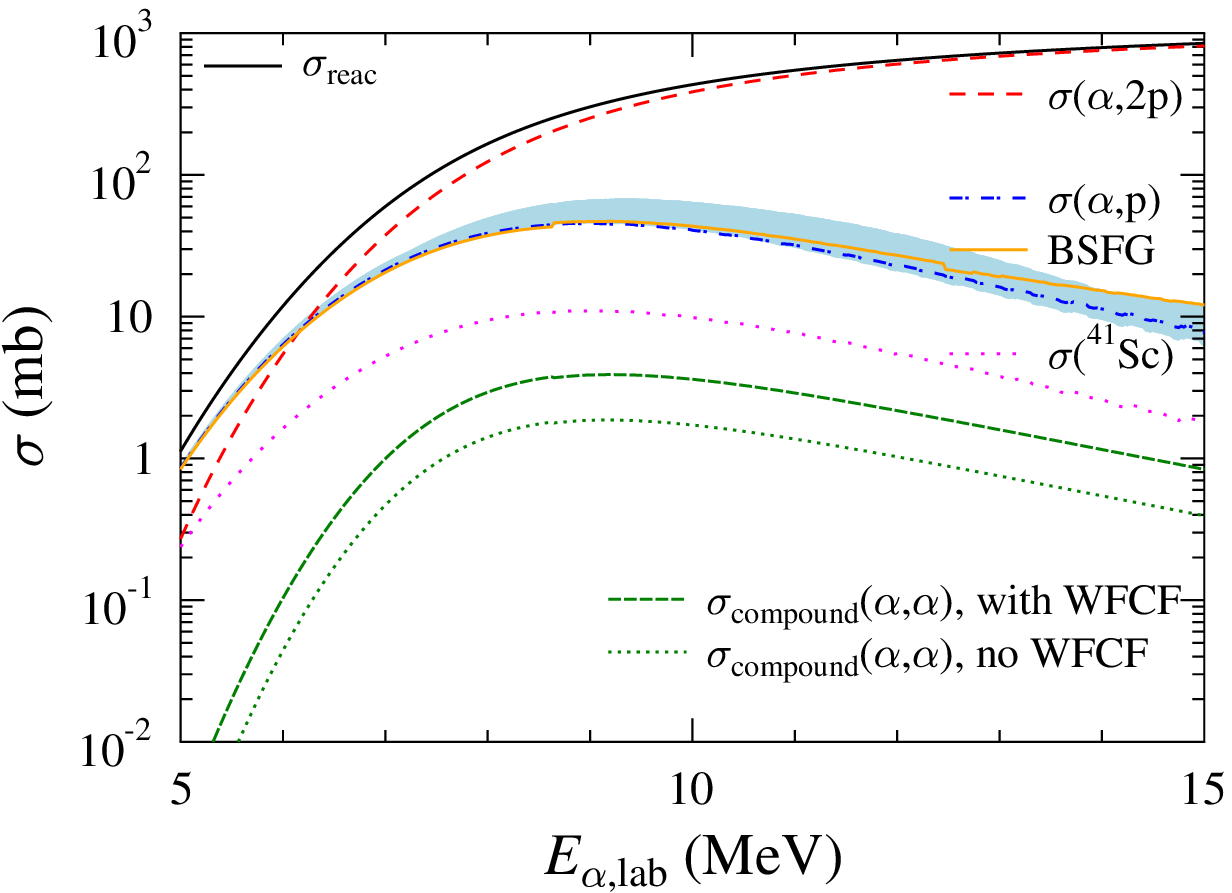}
\caption{
\label{fig:ca38}
(Color online) Total cross section \stot\ (full black line), compound-elastic
$\sigma_{\rm{compound}}$\raa\ (green dashed and dotted), and \caviii \rap
\sci\ (blue dash-dotted) and \caviii \rapp \canull\ (red long-dashed) reaction
cross sections. The uncertainty of the \caviii \rap \sci\ cross section from
the choice of different LDs is indicated by the blue shaded area. The dotted
magenta line indicates the production cross section of \sci , i.e.\ without
contributions from higher-lying levels in the residual \sci\ which decay
preferentially by proton emission. The full orange line for the \rap\ cross
section is calculated from the back-shifted Fermi gas LD and will be discussed
later in Sect.~\ref{sec:ar40}.
}
\end{figure}

The $Q$-values of the \caviii \ran \tii\ and \caviii \rap \sci\ reactions are
$Q_n = -12.01$ MeV and $Q_p = +1.72$ MeV. Thus, proton emission from the
\tiii\ compound nucleus dominates at low energies, and the compound-elastic
channel is much weaker. Consequently, the width fluctuation correction is not
relevant for these reactions. However, the residual nucleus \sci\ of the
\caviii \rap \sci\ reaction has only a very small proton separation energy of
$S_p = 1.09$ MeV, and thus also the \rapp\ channel is open at all
energies. According to TALYS, the \rap\ channel dominates below about 6 MeV
whereas at higher energies the \rapp\ channel exceeds the
\rap\ contribution. The \ran\ cross section remains below 1 mb up to 15
MeV. The results are shown in Fig.~\ref{fig:ca38}.

The calculation of the cross sections for \caviii\ uses the reference
parameters which were fixed for \arviii\ (as explained above). Thus, the
uncertainty of the predicted cross sections should be relatively low. As the
LD was not well-constrained by the \arviii\ data, various parametrizations of
the LD were used to estimate the resulting uncertainty for the \caviii \rap
\sci\ cross section (blue-shaded area in Fig.~\ref{fig:ca38}).

As already mentioned above, the role of the width fluctuation correction
remains minor. Similar to \arviii , the WFCF enhances the compound-elastic
contribution by about a factor of two. However, because the compound-elastic
channel is at least two orders of magnitude below the total cross section
\stot\ for \caviii , the WFCF has practically no influence on the dominating
\rap\ (at low energies) and \rapp\ channels (at higher energies above 6 MeV). 

Because of the astrophysical relevance of \rap\ cross sections of $T_z = -1$
nuclei, the reference parameters are also used to calculate the astrophysical
reaction rate \Nsv\ for the \caviii \rap \sci\ reaction. The results are
listed in Table \ref{tab:ca38rate}.
\begin{table}[tbh]
\caption{\label{tab:ca38rate}
Astrophysical reaction rate \Nsv\ of the \sci\ production from the \caviii
\rap \sci\ reaction, calculated from the \arviii\ reference parameters. The
energy $E_0$ of the classical Gamow window is given to estimate the relevant
energy range for the calculation of \Nsv .  
}
\begin{center}
\begin{tabular}{r@{~~~~~~}rl@{~~~~~~~~~}r}
\hline
\multicolumn{1}{l}{$T_9$}
& \multicolumn{2}{l}{\Nsv }
& \multicolumn{1}{c}{$E_0$} \\
\multicolumn{1}{l}{$-$}
& \multicolumn{2}{l}{(cm$^3$\,s$^{-1}$\,mole$^{-1}$)}
& \multicolumn{1}{c}{(keV)} \\
\hline
 0.1 &  6.74 & $\times 10^{-47}$ &       472 \\
 0.2 &  1.80 & $\times 10^{-32}$ &       749 \\
 0.5 &  3.87 & $\times 10^{-18}$ &      1380 \\
 0.8 &  1.75 & $\times 10^{-12}$ &      1888 \\
 1.0 &  3.87 & $\times 10^{-10}$ &      2191 \\
 1.2 &  2.26 & $\times 10^{-08}$ &      2474 \\
 1.5 &  2.21 & $\times 10^{-06}$ &      2871 \\
 2.0 &  4.44 & $\times 10^{-04}$ &      3478 \\
 2.5 &  1.73 & $\times 10^{-02}$ &      4035 \\
 3.0 &  2.60 & $\times 10^{-01}$ &      4557 \\
\hline
\end{tabular}
\end{center}
\end{table}

The calculation of the astrophysical reaction rate \Nsv\ requires an
additional consideration of the decay properties of all final states in the
residual nucleus \sci . The \sci\ production cross section from the
\rap\ reaction (as provided by TALYS) is composed essentially of the
contributions of several low-lying states in \sci\ which are taken into
account explicitly, see Eq.~(\ref{eq:Ti}). According to the ENSDF database
\cite{ENSDF,NDS}, only one excited state with $J^\pi = 7/2^+$ at $E^\ast =
2882$ keV has a noticeable branching to the $J^\pi = 7/2^-$ ground state of
\sci\ whereas the other excited states in \sci\ decay preferentially by proton
emission. Therefore the \sci\ production cross section in TALYS has to be
corrected accordingly; this leads to a slight reduction of the cross section
in the astrophysically relevant energy region by less than a factor of two and
a strong reduction at higher energies. The \sci\ production cross section is
shown in Fig.~\ref{fig:ca38} as magenta dotted line.

\begin{figure}[htb]
\includegraphics[width=\columnwidth,clip=]{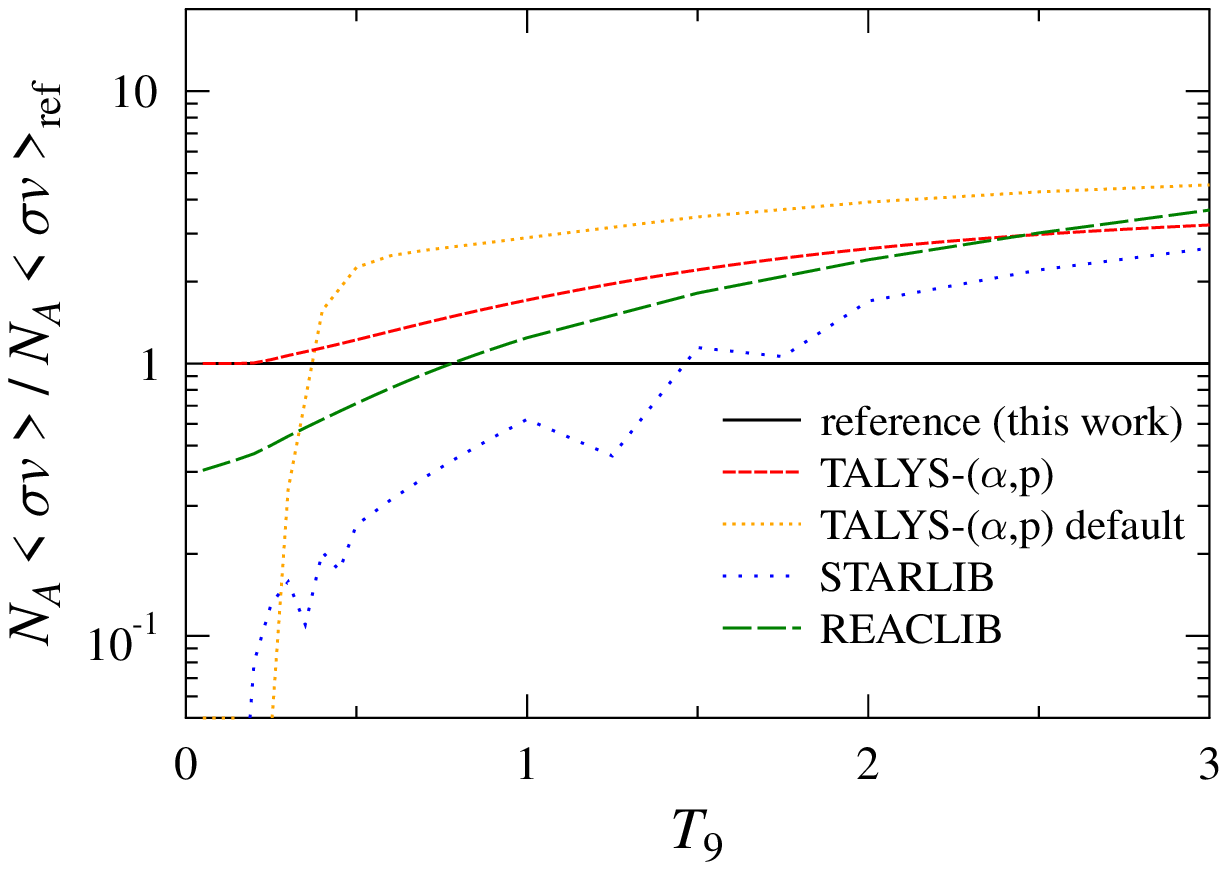}
\caption{
\label{fig:ca38rate}
(Color online)
Astrophysical reaction rate \Nsv\ of the \sci\ production from the \caviii
\rap \sci\ reaction: comparison of the new reference rate from the
\arviii\ reference parameters to various previous calculations
\cite{TALYS,NONSMOKER,Rau00,STARLIB,STARLIB_REF,REACLIB,REACLIB_REF}. 
For better visualization, all rates are normalized to the new reference rate
\Nsv $_{\rm{ref}}$ from this work. Further discussion see text.
}
\end{figure}

The new recommended rate of the \sci\ production from the \caviii \rap
\sci\ reaction is compared to previous calculations in
Fig.~\ref{fig:ca38rate}. Significant discrepancies to all previous evaluations
are found which result from the choice of the A-OMP, from the consideration of
the preferential proton decay of excited states in \sci , and from the
numerical treatment.

The relevance of the proton decay of excited states in \sci\ is illustrated by
the comparison with the rate which is calculated from the \rap\ cross section
as provided by TALYS (red short-dashed line in Fig.~\ref{fig:ca38rate},
labeled ``TALYS-\rap ``). At very low temperatures below $T_9 \approx 0.3$,
the Gamow window is located below 1 MeV. Because of the only slightly positive
$Q$-value of $Q_p = +1.72$ MeV, excited states in the residual \sci\ nucleus
do not contribute in the \caviii \rap \sci\ reaction below $T_9 \approx 0.3$,
and the rate is identical to the reference rate. Above $T_9 = 0.5$, the
contributions of excited states increase and lead to an increased rate by
about a factor of 1.7 at $T_9 = 1$ and more than a factor of 3 at $T_9 =
3$. Thus, the reference rate \Nsv $_{\rm{ref}}$ shows a different temperature
dependence. Note that these rates were calculated numerically from the TALYS
cross sections in small steps of 5 keV to avoid numerical complications at low
temperatures. The numerical stability was checked carefully, see also the
discussion in \cite{Tal18}.

The role of different A-OMPs is illustrated by the TALYS default rate and by
the rate in STARLIB \cite{STARLIB,STARLIB_REF}. The TALYS default rate (orange
dotted, ``TALYS-\rap\ default'' in Fig.~\ref{fig:ca38rate}) is based on the
A-OMP by Watanabe which leads to increased cross sections by about a factor of
two in the astrophysically relevant energy region around $T_9 = 1$. As the
TALYS default calculation does not take into account the proton decay of
excited states in \sci , the influence of the A-OMP can be seen best by
comparison to the ``TALYS-\rap `` curve in Fig.~\ref{fig:ca38rate} which also
neglects the proton decay of excited states in \sci . The strong decrease of
the TALYS default rate at low temperatures below $T_9 \approx 0.3$ results
probably from numerics because TALYS automatically selects about 230 energies
from 0 to 50 MeV to calculate \Nsv . (The TALYS default rate has been
calculated within TALYS whereas the previously discussed rates have been
calculated outside TALYS by numerical integration of TALYS cross sections in
very small steps.)

Contrary to the TALYS default rate, the rate in STARLIB is based on the A-OMP
by Demetriou {\it et al.}\ \cite{Dem02} in its third version
\cite{Gor18}. This A-OMP typically shows lower cross sections than other
A-OMPs at low energies. This trend to lower cross sections becomes also
visible for the \arviii\ mirror nucleus at the lowest energies in
Fig.~\ref{fig:ar38_aomp} but very low energies are not accessible
for \arviii\ because of the negative $Q$-values of the \ran\ and
\rap\ reactions. The lower \rap\ cross sections from the Demetriou {\it et
  al.}\ A-OMP lead to lower reaction rates at low temperatures (blue dotted
line in Fig.~\ref{fig:ca38rate}). At higher temperatures the Demetriou {\it et
  al.}\ rate exceeds the reference rate because proton emission from excited
states in the \sci\ residual nucleus was not taken into account. Similar to
the TALYS default rate, also this rate shows was calculated within TALYS and
shows a similar steep drop towards the lowest temperatures below $T_9 \approx
0.3$.

The REACLIB \cite{REACLIB,REACLIB_REF} rate is taken from the NON-SMOKER
calculations by Rauscher and Thielemann \cite{NONSMOKER,Rau00}. NON-SMOKER
uses the same A-OMP as the reference calculation in the present study, but
does not take into account two-particle emission in the exit channel. This
leads to an overestimation of the rate at higher temperatures because the
dominating \rapp\ channel at higher energies is completely neglected (green
dashed line in Fig.~\ref{fig:ca38rate}). At lower temperatures the REACLIB
rate should approach the reference rate because the same A-OMP was used. The
reason for the deviation by a factor of about 2 below $T_9 \approx 0.5$ is not
clear; it may be related to the fact that REACLIB usually provides rates from
their fit function instead of the underlying calculation.

Summarizing, the production rate \Nsv\ of \sci\ from the \caviii \rap
\sci\ reaction in this work shows a different temperature dependence with
lower rates at high temperatures because of the dominating proton decay of
excited states in the residual \sci\ nucleus. Around $T_9 \approx 1$, the
choice of the reference A-OMP by McFadden/Satchler leads to a rate between the
high rate from the TALYS default potential by Watanabe and the low rate from
the Demetriou potential which was used for STARLIB. Because of these findings,
further investigations of the \rap\ cross sections in the \al p-process along
isospin $T_z = -1$ nuclei are required to provide all \rap\ reaction rates in
a consistent way and to reduce the uncertainties of the calculated
\rap\ reaction rates.

\subsection{$^{36}$Ar}
\label{sec:ar36}
The nucleus \arvi\ is one of the two remaining outliers in the systematics of
\cite{Mohr15}. Unfortunately, there is only one experimental data point for
the \arvi \rap \kix\ reaction which has been measured more than 60 years ago
by Schwartz {\it et al.}\ \cite{Schwartz56}.

\begin{figure}[htb]
\includegraphics[width=\columnwidth,clip=]{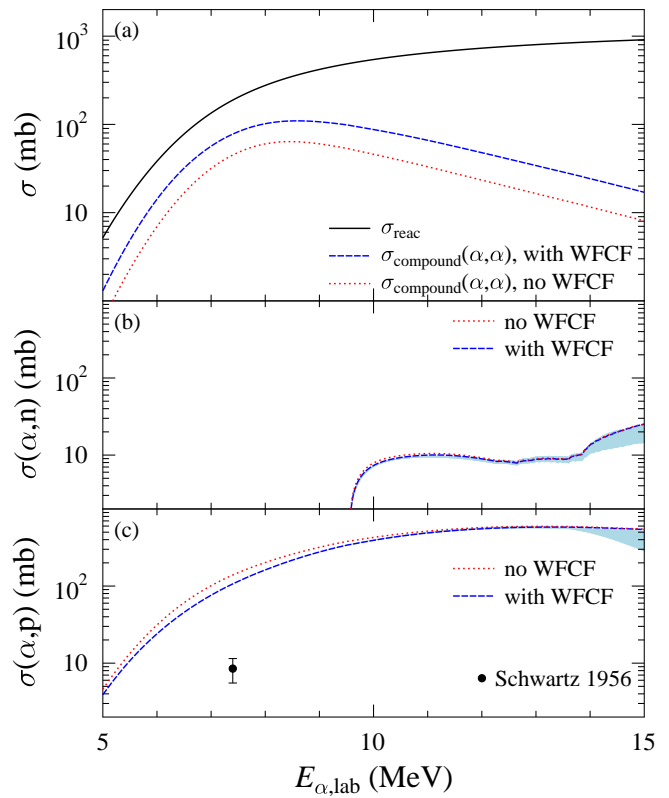}
\caption{
\label{fig:ar36}
(Color online)
Total cross section \stot\ and compound-elastic
$\sigma_{\rm{compound}}$\raa\ for \arvi\ (upper, a), calculated from the
reference parameters without WFCF (dotted red) and with WFCF (dashed
blue). Because of the increased $\sigma_{\rm{compound}}$\raa , the
\ran\ (middle, b) and \rap\ (lower, c) cross sections are slightly reduced by
the WFCF. The minor sensitivity to the choice of the LD is illustrated by the
blue-shaded areas. Further discussion see text. 
}
\end{figure}

The cross sections for \arvi\ have been calculated from the reference
parameters of \arviii\ (as defined above) and are shown in
Fig.~\ref{fig:ar36}. As for the other nuclei under study, the width
fluctuation correction enhances the compound-elastic channel by about a factor
of two. The compound-elastic contribution is smaller than in the
\arviii\ case, but still significant. Consequently, there is a noticeable
reduction of the \arvi \rap \kix\ cross section by about 25\% around the
energy of the experimental point at $E_{\alpha,{\rm{lab}}} = 7.4$
MeV. Although the reduction brings the calculation somewhat closer to the
experimental data point, there still remains a huge discrepancy of at least
one order of magnitude. As already pointed out in \cite{Mohr15}, the energy of
the data point requires a correction by about $-500$ keV because
$E_{\alpha,{\rm{lab}}} = 7.4$ MeV in \cite{Schwartz56} is the nominal beam
energy without corrections for the energy loss in the entrance window of the
target and in the target gas. But even this correction does not lead to
reasonable agreement between the experiment and the present improved
calculation. New experimental data are needed to confirm or to resolve the
discrepancy between experiment on the one hand and calculation and systematics
\cite{Mohr15} on the other hand.

Similar to the \caviii\ case, the uncertainty of the \arvi \rap \kix\ cross
section from the choice of the LD was investigated. It turns out that the
level density in the residual nuclei is quite low, and thus the transmissions
$T_i$ in Eq.~(\ref{eq:Ti}) are essentially defined by the sum over known
low-lying levels in the first term on the l.h.s.\ of
Eq.~(\ref{eq:Ti}). Consequently, the choice of various LDs for the \arvi \rap
\kix\ reaction leads to practically identical cross sections at low energies
(within a line width in Fig.~\ref{fig:ar36} below about 13 MeV). The range of
calculations from different LDs is illustrated by the blue-shaded area in
Fig.~\ref{fig:ar36}. A similar small sensitivity to the choice of the LD is
found for the \arvi \ran \caix\ reaction.

\subsection{$^{40}$Ar}
\label{sec:ar40}
\begin{figure}[htb]
\includegraphics[width=\columnwidth,clip=]{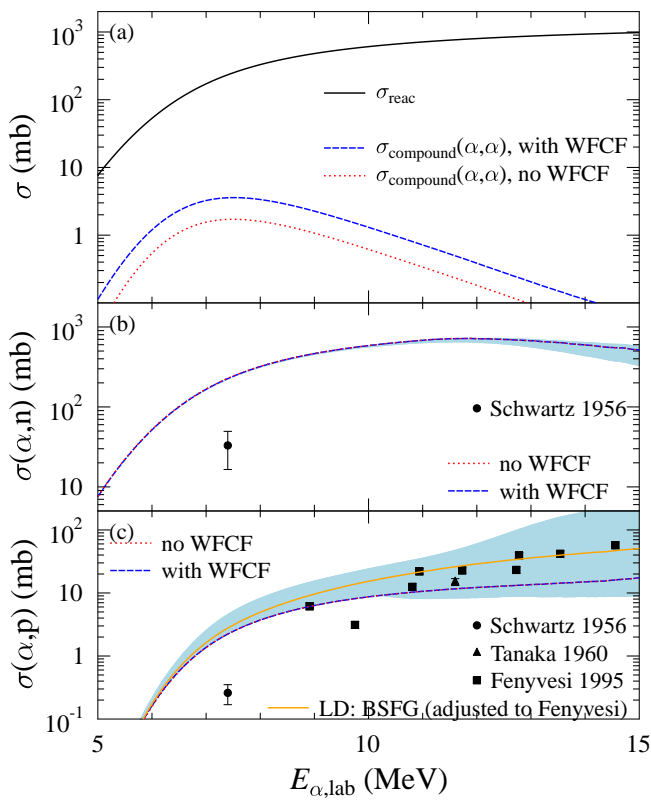}
\caption{
\label{fig:ar40}
(Color online)
Total cross section \stot\ and compound-elastic
$\sigma_{\rm{compound}}$\raa\ for \arnull\ (upper, a), calculated from the
reference parameters without WFCF (dotted red) and with WFCF (dashed
blue). The \ran\ and \rap\ cross sections are shown in the middle (b) and
lower (c) parts. The range of calculations from different parametrizations of
the LD is shown as lightblue shaded area. Whereas the \ran\ data are
relatively insensitive to the choice of the LD, the description of the
\rap\ data can be improved using a different LD. Further discussion see text.
}
\end{figure}

The procedure of the previous Sect.~\ref{sec:ar36} was repeated for
\arnull\ which is the second remaining outlier in the systematics of
\cite{Mohr15}. Again, the choice of the reference parameters in combination
with a proper treatment of the WFCFs should lead to a reliable prediction of
the \arnull \ran \caiii\ and \arnull \rap \kiii\ cross sections. The results
are shown in Fig.~\ref{fig:ar40} and compared to experimental data
\cite{Schwartz56,Tanaka60,Feny95}. The \ran\ and \rap\ reactions have both
slightly negative $Q$-values with $Q_n = -2.28$ MeV and $Q_p = -3.33$ MeV. In
the shown energy range between 5 and 15 MeV, the \ran\ channel dominates and
is very close to the total reaction cross section \stot . Again, the
compound-elastic cross section is enhanced by about a factor of two, but this
enhancement of the weak compound-elastic channel does practically not affect
the stronger \ran\ and \rap\ channels, and thus the calculated reaction cross
sections with and without WFCFs are practically identical.

As for \arvi , the early data by Schwartz {\it et al.}\ \cite{Schwartz56} are
overestimated by almost one order of magnitude, and a significant
overestimation persists also after a correction of the energy by about $-500$
keV (as for \arvi\ as discussed in the previous Sect.~\ref{sec:ar36} and in
\cite{Mohr15}). 

Interestingly, the later \rap\ data by Tanaka {\it et al.}\ \cite{Tanaka60}
(only one data point below 15 MeV) and by Fenyvesi {\it et al.}\ \cite{Feny95}
were also overestimated by the calculation in \cite{Mohr15}. However, the
present detailed study shows that the calculated \arnull \rap \kiii\ cross
section at higher energies depends sensitively on the chosen parametrization
of the LD whereas the dominating \arnull \ran \caiii\ cross section is only
weakly affected. The range of calculated cross sections from different LDs is
indicated in Fig.~\ref{fig:ar40} as shaded area. The TALYS default level
density which is based on the constant temperature plus Fermi gas model (as
used in \cite{Mohr15}) leads to an overestimation of the experimental
\rap\ data. Contrary to this, the reference LD from the generalized superfluid
model slightly underestimates the experimental data, and the LD from the
back-shifted Fermi gas (BSFG) model reproduces the experimental data
well. Thus, it seems that at least for the residual odd-even nuclei with $A =
43$ the BSFG LD is a better choice than the reference LD from the generalized
superfluid model.

Among the cross sections of the other nuclei \arvi , \arviii , and \caviii ,
only the \caviii \rap \sci\ cross section is slightly sensitive to the chosen
LD. Therefore, those calculations were repeated using the BSFG LD instead of
the reference LD from the generalized superfluid model. Below 10 MeV the
calculated result from the BSFG LD is practically identical to the calculation
with the reference LD, and above 10 MeV the BSFG LD leads to slightly higher
cross sections (see orange line in Fig.~\ref{fig:ca38}).

If the final calculation of the \arnull \rap \kiii\ cross section is
considered as reliable in the entire energy range of Fig.~\ref{fig:ar40}, a
correction factor for the early data by Schwartz {\it et
  al.}\ \cite{Schwartz56} can be derived. Assuming a 500 keV shift to lower
energies because of the energy loss in the entrance window (as suggested in
\cite{Mohr15}), a correction factor of about 5 is found. The same correction
procedure (500 keV energy shift and increase of the cross section by a factor
of five) leads also to good agreement between the calculations and the
experiment of Schwartz {\it et al.} \cite{Schwartz56} for the \arnull \ran
\caiii\ and \arvi \rap \kix\ reactions. This finding can be considered as
evidence that such a correction is indeed required for the early Schwartz {\it
  et al.}\ data.

\section{Isospin considerations}
\label{sec:iso}
As pointed out in Sec.~\ref{sec:sens}, the successful application of the StM
is based on a sufficiently high level density in the compound nucleus, and
thus the StM is applicable for intermediate mass and heavy nuclei. Under these
circumstances the role of isospin conservation is very minor, and typical
computer codes like TALYS do not consider isospin explicitly for the
calculation of the transmission coefficients in Eq.~(\ref{eq:Ti}). However,
for light nuclei, in particular for target nuclei with isospin $T = 0$ ($N =
Z$), substantial changes may occur for \al -induced reactions. These changes
are related to the fact that the \al\ projectile with $T = 0$ can only
populate states in the compound nucleus with the same isospin $T$ as the
target nucleus. Without explicit consideration of isospin, Eq.~(\ref{eq:Ti})
overestimates the transmissions to the proton and neutron channels because of
additional isospin couplings which result from $T \ne 0$ of the
ejectiles. Contrary, the \al\ channel with $T = 0$ is calculated correctly in
Eq.~(\ref{eq:Ti}). The resulting suppression of the \ran\ and \rap\ channels
has been discussed in detail by Grimes \cite{Gri92}, and Table I of
\cite{Gri92} shows the isospin couplings for \al -induced reactions on targets
with isospin $T = 0, 1/2$, and $1$. We follow the idea of \cite{Gri92},
starting with the $T = 2$ target \arnull , and provide an approximate
correction for the TALYS cross sections. Detailed information on the role of
isospin in StM calculations for nuclear astrophysics is also provided in
\cite{Rau98,Rau00b}; in particular, \cite{Rau00b} focuses on the important
point of isospin suppression in \rag\ capture reactions in self-conjugate $N =
Z$ nuclei.

\subsection{Target \arnull , compound \caiv : $T = 2$, $T_z = +2$}
\label{sec:ar40iso}
According to \cite{Gri92}, the isospin coupling is given by the square of the
respective Clebsch-Gordan coefficient which couples the isospins $T_E$ of the
ejectile and $T_R$ of the residual nucleus to the isospin $T_C$ of the
compound nucleus. For the \arnull\ target we find for the \al\ channel $< T_R
\, T_{z,R} \, \, T_E \, T_{z,E} \, | \, T_C \, T_{z,C} > = 1.0$; the coupling
to the \al\ channel is 1.0 (this result also holds for all reactions under
study). For the proton channel with the residual \kiii\ ($T_R = 5/2$, $T_{z,R}
= +5/2$) we obtain a Clebsch-Gordan coefficient of $\sqrt{5/6}$, leading to a
coupling of 5/6. For the neutron channel with the residual \caiii\ ($T_R =
3/2$, $T_{z,R} = +3/2$) the coupling is 1.0; contributions of higher-lying
states in \caiii\ with $T_R = 5/2$; $T_{z,R} = +3/2$ with a coupling of 1/6
are neglected. The given couplings extend Table I of \cite{Gri92} for the case
of $T_C = 2$, $T_{z,C} = +2$. 

An approximate correction to the TALYS calculations in the previous sections
can be made as follows. The correction is based on the assumptions of isospin
conservation (also excluding isospin mixing) and the independence of the
transmission coefficients on the isospin. It will be shown that the resulting
corrections are minor for the reactions under study, and thus the above
simplifying assumptions have no major effect on the final conclusions of the
present study. Furthermore, as isospin conservation is violated to some
extent, the following correction may be considered as an upper limit for the
relevance of isospin in the StM.

The isospin-corrected cross sections $\sigma_{\rm{iso}}(\alpha,X)$ are given by
\begin{equation}
\sigma_{\rm{iso}}(\alpha,X) = {\cal{N}}\,  w_X \, \sigma(\alpha,X)
\label{eq:iso}
\end{equation}
where the $w_X$ are the isospin couplings (as provided above and in Table I of
\cite{Gri92}), and ${\cal{N}}$ is a normalization factor to fulfill
\begin{eqnarray}
\sigma_{\rm{reac}} 
   & \approx & 
   \sigma(\alpha,{\rm{n}}) + 
   \sigma(\alpha,{\rm{p}}) + 
   \sigma(\alpha,\alpha) \nonumber \\
   & \approx & 
   \sigma_{\rm{iso}}(\alpha,{\rm{n}}) + 
   \sigma_{\rm{iso}}(\alpha,{\rm{p}}) + 
   \sigma_{\rm{iso}}(\alpha,\alpha)
   \label{eq:sum_iso}
\end{eqnarray}
for the total \al -induced reaction cross section \stot\ at low energies;
other open channels like e.g.\ \rag\ are typically weak and are neglected in
Eq.~(\ref{eq:sum_iso}). For \arnull\ this results in 
\begin{equation}
{\cal{N}} = 
  \biggl[ 1 - \frac{\sigma(\alpha,{\rm{p}})}{6 \, \sigma_{\rm{reac}}} \biggr]^{-1}
\label{eq:norm_ar40}
\end{equation}
from $w_{\rm{p}} = 5/6$ and $w_{\rm{n}} = w_\alpha = 1$. Note that ${\cal{N}}
\ge 1$ close to unity, thus leading to an enhancement of the reaction channels
with $w_X = 1$ and to a reduction for channels with $w_X < 1$. Because the
\rap\ contribution to \stot\ does not exceed a few per cent in the energy
range under study (see Fig.~\ref{fig:ar40}), the normalization factor
${\cal{N}}$ in Eq.~(\ref{eq:norm_ar40}) remains very close to unity for
\arnull . According to Eq.~(\ref{eq:iso}), the \rap\ cross section is thus
reduced by about a factor of 5/6 which is inside the shown uncertainty from
the choice of the level density, and the other channels are practically not
affected by the isospin correction.

\subsection{Target \arviii , compound \caii : $T = 1$, $T_z = +1$}
\label{sec:ar38iso}
The respective numbers for the couplings are $w_{\rm{p}} = 3/4$ and
$w_{\rm{n}} = w_\alpha = 1$, and the normalization is given by
\begin{equation}
{\cal{N}} = 
  \biggl[ 1 - \frac{\sigma(\alpha,{\rm{p}})}{4 \, \sigma_{\rm{reac}}} \biggr]^{-1}
\label{eq:norm_ar38}
\end{equation}
As the \rap\ contribution remains below 40\% for the whole energy range under
study, the normalization factor does not exceed ${\cal{N}} \approx 1.1$,
leading only to a slight enhancement of the \ran\ and \raa\ channels and a
reduction of the \rap\ cross section by about 20\%.

In Sec.~\ref{sec:NOMP} and Fig.~\ref{fig:ar38_nomp} it was pointed out that
the branching between the \ran\ and \rap\ channels depends essentially on the
chosen N-OMP, and it was concluded that only the TALYS default N-OMP by Koning
and Delaroche \cite{Kon03} is able to reproduce the new experimental data
\cite{Tal18}. The JLM-type potentials showed a trend to overestimate the
\rap\ cross sections and underestimate the \ran\ cross sections. These
deviations of the JLM-type potentials are approximately compensated by the
isospin corrections from Eq.~(\ref{eq:iso}), and thus the clear preference for
the TALYS default potential by Koning and Delaroche \cite{Kon03} is weakened
by the isospin correction in Eq.~(\ref{eq:iso}).

\subsection{Target \caviii , compound \tiii : $T = 1$, $T_z = -1$}
\label{sec:ca38iso}
Here the results for $w_X$ and ${\cal{N}}$ can be taken from the previous
Sec.~\ref{sec:ar38iso}; only the role of neutrons and protons has to be
exchanged in the given numbers for ${\cal{N}}$ and $w_X$. Because of the
strongly negative $Q$-value of $Q_n \approx -12$ MeV, the isospin corrections
vanish for almost the full energy range under study. In particular, the
calculated reaction rates in Table \ref{tab:ca38rate} are not affected by the
isospin correction.

\subsection{Target \arvi , compound \canull : $T = 0$, $T_z = 0$}
\label{sec:ar36iso}
As already pointed out in \cite{Gri92}, the largest corrections are expected
for \al -induced reactions on $N = Z$ target nuclei with $T = 0$. Here we find
$w_{\rm{n}} = w_{\rm{p}} = 1/2$, $w_\alpha = 1.0$, and
\begin{equation}
{\cal{N}} = 
  \biggl[ 1 - \frac{\sigma(\alpha,{\rm{n}}) + \sigma(\alpha,{\rm{p}})}{2 \,
      \sigma_{\rm{reac}}} \biggr]^{-1} 
\label{eq:norm_ar36}
\end{equation}
which leads to an enhancement of the \al\ channel and to a reduction of the
\ran\ and \rap\ cross sections. However, the reduction does not reach a factor
of two (as one might expect from the coupling of 1/2 for the neutron and
proton channels) because the total reaction cross section is dominated by the
\rap\ contribution. Note that in the extreme case of \stot\ $\approx$
$\sigma$\rap , the normalization approaches ${\cal{N}} \approx 2$, thus
compensating the reduction by the coupling of 1/2 in Eq.~(\ref{eq:iso}); in
this case the strong enhancement of the tiny \raa\ contribution by a factor of
two does practically not affect the dominating \rap\ channel.

\begin{figure}[htb]
\includegraphics[width=\columnwidth,clip=]{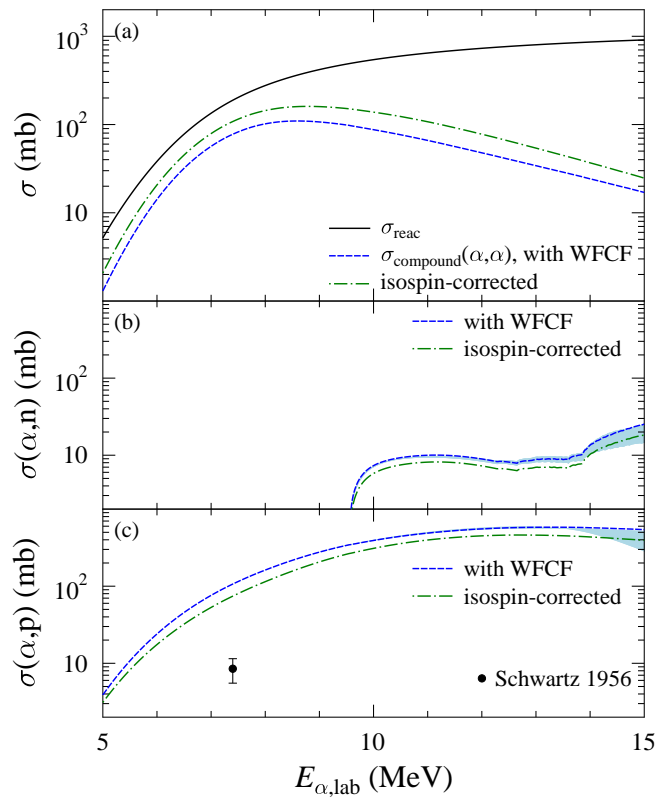}
\caption{
\label{fig:ar36iso}
(Color online)
Same as Fig.~\ref{fig:ar36}, but with isospin correction: Total cross section
\stot\ and compound-elastic $\sigma_{\rm{compound}}$\raa\ for \arvi\ (upper,
a), calculated from the reference parameters with width fluctuation correction
(dashed blue) and with additional isospin correction (dash-dotted green) from
Eq.~(\ref{eq:iso}). The isospin correction leads to an increased
$\sigma_{\rm{compound}}$\raa , and it reduces the \ran\ (middle, b) and
\rap\ (lower, c) cross sections by about 30\%. 
}
\end{figure}

The isospin-corrected cross sections for \arvi\ are shown in
Fig.~\ref{fig:ar36iso}. It is obvious from Fig.~\ref{fig:ar36iso} that the
reduction of the \arvi \rap \kix\ cross section from the isospin correction in
Eq.~(\ref{eq:iso}) is only about 30\% at 7.5 MeV and thus cannot resolve the
huge discrepancy to the experimental data point by Schwartz {\it et
  al.}\ \cite{Schwartz56}.

\section{Conclusions}
\label{sec:conc}
The new experimental data for the \arviii \ran \cai\ and \arviii \rap
\ki\ reactions \cite{Tal18} were analyzed within the statistical model using
the TALYS code. Best-fit input parameters for the statistical model
calculations were determined by a $\chi^2$-based assessment \cite{Tal18}. The
present study provides a careful discussion of the uncertainties from the
different ingredients of the statistical model. It is found that a very good
description of the experimental data can only be achieved from the
$\alpha$-nucleus potential by McFadden and Satchler \cite{McF66} in
combination with the default nucleon potential by Koning and Delaroche
\cite{Kon03}. The smallest $\chi^2$ is furthermore achieved for the level
density from the generalized superfluid model, but because of the minor
sensitivity to the level density, other parametrizations of the level density
are not excluded by the new data. As the experimental \ran\ and \rap\ data for
\arviii\ are not sensitive to the chosen $\gamma$-ray strength, no conclusion
can be drawn on the choice of the $\gamma$-ray strength function. In addition,
the importance of a proper treatment of the width fluctuation correction is
pointed out especially for the \arviii \ran \cai\ and \arviii \rap
\ki\ reactions. Isospin corrections to the calculated reaction cross sections
reduce the \arvi \ran \caix\ and \arvi \rap \kix\ cross sections by about
30\%, play a minor role with about $10 - 20$\% correction for \arviii , and
are practically negligible for \caviii\ and \arnull .

The best-fit parameters from the \arviii\ data are used as reference
parameters to predict \al -induced cross sections for the isospin mirror
nucleus \caviii\ and for the neighboring argon isotopes \arvi\ and
\arnull\ with improved reliability, and a new astrophysical reaction rate
\Nsv\ is calculated for the \caviii \rap \sci\ reaction. This new \Nsv\ shows
a different temperature dependence than previous calculations because proton
decay from excited states in the residual \sci\ nucleus was not taken into
account in earlier work. However, before firm conclusions on astrophysical
consequences can be drawn, further \rap\ reactions on the isospin $T_z = -1$
nuclei from $^{22}$Mg to $^{34}$Ar should also be re-investigated.

For \arvi\ the disagreement between the early experimental data by Schwartz
{\it et al.}\ \cite{Schwartz56} for the \arvi \rap \kix\ reaction persists,
and this holds also for the \arnull \ran \caiii\ and \arnull \rap
\kiii\ reactions. But it was found that more recent data for the \arnull \rap
\kiii\ reaction \cite{Tanaka60,Feny95} can now be reproduced in a calculation
which uses the well-constrained reference parameters for the \al -nucleus
potential and the nucleon-nucleus potential and a different level density from
the back-shifted Fermi gas model. \arvi\ and \arnull\ have been identified as
the two remaining outliers in the general systematics of \al -induced cross
sections in the $A \approx 20 - 50$ mass range \cite{Mohr15}. At least for
\arnull , the role as outlier is reduced by the improved reproduction of
\arnull \rap \kiii\ data of \cite{Tanaka60,Feny95}. In combination with the
good description of the new data for \arviii , the \al -induced cross sections
for \arvi\ and \arnull\ may also be considered as regular if all early data by
Schwartz {\it et al.}\ \cite{Schwartz56} for \arvi\ and \arnull\ are shifted
by about 500 keV to lower energies and increased by about a factor of five.

\acknowledgments 
We thank Ernst Rehm for encouraging discussions, Stephane Goriely and
Christian Iliadis for additional information on the STARLIB database, Arjan
Koning for his help with TALYS, and the anonymous referee for pointing to the
relevance of isospin in statistical model calculations. This work was
supported by NKFIH (K108459 and K120666) and U.S. Department of Energy, Office
of Science, Office of Nuclear Physics, under Contract No.\ DE-AC02-06CH11357.

\end{document}